\documentclass[11pt, letter,fleqn]{article}
\usepackage{amsmath, amssymb,amsthm,cite}

\title{Computation of fluxes of conservation laws}
\author{

Alexei F. Cheviakov\footnotemark[1], \\
{\small \emph{Department of Mathematics and Statistics, University of Saskatchewan, Saskatoon, S7N 5E6 Canada}}}

\newtheorem{theorem}{Theorem}

\newtheorem{corollary}{Corollary}

\def\const{\hbox{\rm const}}

\def\grad{\mathop{\hbox{\rm grad}}}

\def\max{\mathop{\hbox{\rm max}}}

\def\div{\mathop{\hbox{\rm div}}}
\def\curl{\mathop{\hbox{\rm curl}}}
\def\vec#1{{\boldsymbol{\rm #1}}}  

\def\PDEs#1#2#3{{\boldsymbol{\rm #1}}\{#2\,; #3\} }
\def\sg#1{{\rm #1}}
\def\beq{\begin{equation}}
\def\eeq{\end{equation}}
\def\barr{\begin{array}{ll}}
\def\earr{\end{array}}

\newcommand{\bfvec}[1]{\mbox{\boldmath{$#1$}}}

{\theoremstyle{definition}
}

{\theoremstyle{definition}
\newtheorem{definition}{Definition}}

{\theoremstyle{definition}
}

{\theoremstyle{definition}
\newtheorem{remark}{Remark}}

\begin{document}

\def \zeroall {
\setcounter{equation}{0} }

\renewcommand{\thedefinition}{\arabic{definition}}
\renewcommand{\theremark}{\arabic{remark}}
\renewcommand{\thetheorem}{\arabic{theorem}}
\renewcommand{\thecorollary}{\arabic{corollary}}

\footnotetext[1]{Electronic mail: cheviakov@math.usask.ca}

\date{}

\maketitle
\begin{abstract}
The direct method of construction of local conservation laws of partial differential equations (PDE) is a
systematic method applicable to a wide class of PDE systems [Anco S. and Bluman G., Direct construction
method for conservation laws of partial differential equations Part II: General treatment. {\sl Europ. J.
Appl. Math.} {\bf 13}, 567--585 (2002)]. Within the direct method, one seeks multipliers, such that the
linear combination of PDEs of a given system with these multipliers yields a divergence expression.

After local conservation law multipliers are found, one needs to reconstruct the fluxes of the conservation law. In this
review paper, we discuss common methods of flux computation, compare them, and illustrate by examples. An implementation of
these methods in symbolic software is also presented.
\end{abstract}




\renewcommand{\baselinestretch}{1.18}\small\normalsize
\renewcommand{\theequation}{\arabic{section}.\arabic{equation}}

\section{Introduction}
\zeroall


Local conservation laws of systems of partial differential equations (PDE), i.e., divergence expressions vanishing on
solutions of a given PDE system, arise in a wide variety of applications and contexts. They often serve as mathematical
expressions for fundamental physical principles, such as conservation of mass, momentum, charge, and energy. Conservation
laws of time-dependent bounded systems describe conserved quantities. Conservation laws of ordinary differential equations
(ODE) yield first integrals.

Conservation laws are widely used in PDE analysis, in particular, studies of existence, uniqueness and stability of
solutions of nonlinear PDEs (e.g., \cite{Lax, Benj, KnopsS}).

Another important area of application of conservation laws is the development and use of numerical methods.
In particular, knowledge of conservation laws of a PDE system of interest enables one to draw  from a wide
variety of robust and efficient numerical schemes which have been derived in the last   decades, including
finite volume, finite element, and discontinuous Galerkin methods (see, e.g., [4, Chapter II; 5, Chapter IV]).

In addition, local conservation laws of a given PDE system are used to introduce nonlocal (potential) variables and thus
construct PDE systems nonlocally related to a given one, with the same solution set. The related framework of nonlocally
related systems (e.g., \cite{bk1987, bkr1988, AGI, SM, BC1, BCI}) has yielded many new results over the recent years.

When a PDE system has an infinite set of local conservation laws involving arbitrary functions, it can be
sometimes mapped into a linear PDE system by an invertible transformation, which can be derived explicitly
when it exists \cite{b_a_wolf}. An infinite countable set of local conservation laws is often associated with
integrability.

In the current paper, we consider local divergence-type conservation laws. Other types of conservation laws (such as
lower-degree, e.g., curl-type, and nonlocal conservation laws) also arise in applications.

We now turn to the question of derivation of conservation laws. Many classical conservation laws have been
found by \emph{ad hoc} methods. One of the most well-known systematic methods is due to Emmy Noether, who
demonstrated that for self-adjoint (variational) PDE systems, conservation laws arise from variational
symmetries, i.e., symmetries that preserve the action integral \cite{Noether}. However, the applicability of
Noether theorem is severely limited, since the majority of PDE systems arising in applications are not
self-adjoint \cite{BlumanKiev05}.

An interesting review devoted to comparison of several methods of computation of conservation laws is found in \cite{t_wolf}.

A more general systematic method of constructing local conservation laws called \emph{the direct method} was suggested in
\cite{AB97,AB02} (see also \cite{BCA2009}). We review this method and the underlying theory in Section \ref{sec:2} of the current
paper. Within this method, one seeks a set of local \emph{multipliers} (also called \emph{factors} or \emph{characteristics})
depending on independent and dependent variables of a given PDE system and derivatives of dependent variables up to some fixed
order, such that a linear combination of the PDEs of the system system taken with these multipliers yields a divergence
expression. Families of multipliers that yield conservation laws are found from determining equations that follow from Euler
differential operators. After finding sets of local conservation law multipliers, one needs to derive expressions for the
corresponding conservation law fluxes.

The goal of this paper is to review and illustrate available methods of flux construction (Section \ref{sec:3}). We start
from a direct (``brute force") method (Section \ref{subsec:meth:direct}) that converts a conservation law directly into the
set of determining equations (linear PDEs) for the unknown fluxes. This method is easy to apply in the case of simple
conservation laws, including those involving arbitrary functions.

The second and the third methods (Sections \ref{subsec:meth:herem} and \ref{subsec:meth:BA})  employ integral formulas,
related to homotopy operators, to compute the fluxes. These formulas can be used to produce closed-form flux expressions for
rather complicated forms of conservation laws, except for the cases when the conservation law involves arbitrary functions.
The first integral formula is presented following \cite{hereman1}. The second integral formula initially appeared in
\cite{AB02}, but is re-derived in the present paper in a different form, which is symmetric with respect to independent
variables, and is simpler for practical computations and software implementation.

The fourth method \cite{AncoScal} (Section \ref{subsec:meth:Anco}) applies to scaling-homogeneous conservation laws of
scaling-invariant PDE systems, which often occur in applications. Within this method, one finds fluxes of conservation laws
through a rather simple formula involving no integration.

Another appropriate reference for the above four methods is [21, Section 1.3].

In Section \ref{sec:5}, we briefly compare and discuss the four flux computation methods presented in the
current paper.

Finally, in Section \ref{sec:6}, we present a symbolic software implementation of the direct method of
conservation law construction and the four flux computation methods in the symbolic package \verb"GeM" for
\verb"Maple" \cite{GeM}, and illustrate these methods with examples.

There also exist other symbolic software packages implementing the direct method, as well as other methods
for computation of conservation laws; see \cite{hereman1, t_wolf2, hereman_soft, Deconinck1, Deconinck2} and
references therein.

\section{Notation and definitions}\label{sec:Notat}
\zeroall

\subsection*{General definitions}

Consider a general system of $N$ partial differential equations of order $k$ with $n$ independent variables $x=(x^1,\ldots
,x^n)$ and $m$ dependent variables $u(x)=(u^1(x),\ldots ,u^m(x))$, denoted by
\beq\label{eq:ch1:secCL:PDEsys}
R^\sigma[u]=R^\sigma(x,u,\partial u, \ldots , \partial^k u)=0, \quad \sigma=1,\ldots ,N.
\eeq
Here partial derivatives are denoted by $u^\mu_i=\frac{\partial u^\mu(x)}{\partial x^i}$;
\[
\partial u \equiv \partial^1 u = \Big( u^1_1(x),\ldots ,u^1_n(x),\ldots ,u^m_1(x), \ldots ,u^m_n(x)\Big)
\]
denotes the set of all first-order partial derivatives;
\[
\barr
\partial^p u&=\left\{ u^\mu_{{i_1} \ldots {i_p}}~|~~\mu=1,\ldots ,m;~~i_1,\ldots ,i_p = 1,\ldots ,n \right\}\\[3ex]
&=\left\{ \dfrac{\partial^p u^\mu(x)}{\partial x^{i_1}\ldots \partial x^{i_p}}~|~\mu=1,\ldots ,m; i_1,\ldots ,i_p = 1,\ldots ,n
\right\}
\earr
\]
denotes the set of all partial derivatives of order $p$.

Within the current paper, the notation ``$f[u]$" means that $f$ is a function of one or more independent variables $x$,
dependent variables $u$, and possibly derivatives of dependent variables, up to some fixed order, i.e.,
\[
f[u] \equiv f(x,u,\partial u, \ldots , \partial^l u),\quad l\geq 0.
\]

\begin{definition}\label{def:solved}
The given PDE system \eqref{eq:ch1:secCL:PDEsys} is written \emph{in a solved form} with respect to some \emph{leading
derivatives}, if
\beq\label{eq:ch1:secCL:PDEsys:solved_lead}
R^\sigma[u]=u^{j_\sigma}_{i_{\sigma,1} \ldots i_{\sigma,s}} - G^\sigma(x,u,\partial u, \ldots ,
\partial^k u)=0, \quad \sigma=1,\ldots ,N,
\eeq
where $s\leq k$, $1\leq j_\sigma\leq m$, $1\leq i_{\sigma,1}, \ldots, i_{\sigma,s}\leq n$ for all $\sigma=1,\ldots,N$. In
\eqref{eq:ch1:secCL:PDEsys:solved_lead}, $\{u^{j_\sigma}_{i_{\sigma,1} \ldots i_{\sigma,s}}\}$ is a set of $N$ linearly
independent $s$th order leading partial derivatives, with the property that none of them or their differential consequences
appears in $\{G^\sigma[u]\}_{\sigma=1}^N$. (For example, for dynamical PDE systems, time derivatives yield a natural set of
leading derivatives.)
\end{definition}

\begin{definition}\label{def:CK}
A PDE system  \eqref{eq:ch1:secCL:PDEsys} \emph{is in Cauchy-Kovalevskaya form} with respect to an independent variable
$x^j$, if the system is in the solved form for the highest derivative of each dependent variable with respect to $x^j$,
i.e.,
\beq\label{eq:ch1:secCL:PDEsys:CKform}
\frac{\partial^{s_{(\sigma)}}}{\partial (x^j)^{s_{(\sigma)}}}u^\sigma=G^\sigma(x,u,\partial u, \ldots ,
\partial^k u), \quad 1\leq s_{(\sigma)} \leq k, \quad \sigma=1,\ldots ,m,
\eeq
where all derivatives with respect to $x^j$ appearing in the right-hand side of each PDE of
\eqref{eq:ch1:secCL:PDEsys:CKform} are of lower order than those appearing on the left-hand side.
\end{definition}

\begin{definition}\label{def:admCK}
A PDE system \eqref{eq:ch1:secCL:PDEsys} \emph{admits a Cauchy-Kovalevskaya form} if it can be written in
Cauchy-Kovalevskaya form \eqref{eq:ch1:secCL:PDEsys:CKform} with respect to some independent variable (after a point
(contact) transformation if necessary).
\end{definition}
A PDE system can admit a Cauchy-Kovalevskaya form only if its number of dependent variables equals the number of PDEs in the
system, i.e., $N = m$. Many PDE systems arising in applications admit a Cauchy-Kovalevskaya form with respect to one or more
of its independent variables. PDE systems that \emph{do not} admit a Cauchy-Kovalevskaya form are typically PDE systems with
differential constraints, such as Maxwell's equations.

\begin{definition}\label{def:TotDer}
The \emph{total derivative operators} with respect to independent variables are given by
\beq\label{eq:ch1:sec11:tot_der}
\sg{D}_{i} = \frac{\partial}{\partial x^i} + u^\mu_{i} \frac{\partial}{\partial u^\mu} + u^\mu_{i i_1}
\frac{\partial}{\partial u^\mu_{{i_1}}} + u^\mu_{i i_1 i_2} \frac{\partial}{\partial u^\mu_{i_1 i_2}}+\ldots,\quad
i=1,\ldots ,n.
\eeq
In \eqref{eq:ch1:sec11:tot_der} and below, summation over repeated indices is assumed.
\end{definition}

\begin{definition}\label{def:EulOp}
The differential \emph{Euler operator} with respect to a function $U(x)$ is defined by
\begin{equation}
\label{eq:ch1:sec13:Bluman:EulerOp} \sg{E}_{U} = \frac{\partial }{\partial U} - \sg{D}_i \frac{\partial }{\partial U_i } +
\ldots + ( - 1)^s\sg{D}_{i_1 } \ldots \sg{D}_{i_s } \frac{\partial }{\partial U_{i_1 \ldots i_s } } + \ldots \;.
\end{equation}
\end{definition}

\subsection*{Local conservation laws}

\begin{definition}
A \emph{local conservation law} of PDE system \eqref{eq:ch1:secCL:PDEsys} is a divergence expression
\beq\label{eq:ch1:secCL:def_CL_nvar}
\div {\bfvec \Phi}[u]\equiv \sg{D}_i \Phi^i[u] \equiv \sg{D}_{1}\Phi^{1}[u]+\ldots + \sg{D}_{n}\Phi^{n}[u]=0
\eeq
holding for all solutions of PDE system \eqref{eq:ch1:secCL:PDEsys}. In \eqref{eq:ch1:secCL:def_CL_nvar}, $\Phi^{i}[u] =
\Phi^{i}(x,u,\partial u, \ldots ,\partial ^{r} u)$, $i=1,\ldots , n$ are called the \emph{fluxes} of the conservation law,
and the highest order derivative ($r$) present in the fluxes $\Phi^i[u]$ is called the (differential) \emph{order of a
conservation law}.
\end{definition}

If one of the independent variables of the PDE system \eqref{eq:ch1:secCL:def_CL_nvar} is time $t$, the conservation law
\eqref{eq:ch1:secCL:def_CL_nvar} takes the form
\beq\label{eq:ch1:secCL:def_CL_nvarT}
\sg{D}_t \Psi[u] + \div\vec{\Phi}[u] = 0,
\eeq
where $\div \vec{\Phi}[u] =\sg{D}_i \Phi^i[u]= \sg{D}_{1}\Phi^{1}[u] +\ldots + \sg{D}_{n-1}\Phi^{n-1}[u]$ is a spatial
divergence, and $x=(x^1, \ldots , x^{n-1})$ are ${n-1}$ spatial variables. Here $\Psi[u]$ is referred to as \emph{a
density}, and $\Phi^i[u]$ as \emph{spatial fluxes} of the conservation law \eqref{eq:ch1:secCL:def_CL_nvarT}.

\begin{definition}\label{def:2}
A local conservation law \eqref{eq:ch1:secCL:def_CL_nvar} of the PDE system \eqref{eq:ch1:secCL:PDEsys} is \emph{trivial} if its
fluxes are of the form $\Phi^i[u]=M^i[u]+H^i[u]$, where $M^i[u]$ vanishes on the solutions of the system
\eqref{eq:ch1:secCL:PDEsys}, and $\sg{D}_{i}H^i[u]\equiv 0$ is a trivial divergence.
\end{definition}

\begin{definition}\label{def:eq}
Two local conservation laws \eqref{eq:ch1:secCL:PDEsys} are \emph{equivalent} if they differ by a trivial
conservation law. Similarly, a set of local conservation laws of the form \eqref{eq:ch1:secCL:PDEsys} is
\emph{linearly dependent} if its nontrivial linear combination (i.e., a linear combination with coefficients
which are not all equal to zero) yields a trivial conservation law.
\end{definition}

\section{The direct method of construction of conservation laws}\label{sec:2}
\zeroall

For a given PDE system, one is interested in finding its nontrivial linearly independent local conservation laws.

We now outline a systematic and generally applicable method of construction of local conservation laws of PDE systems,
called \emph{the direct method} \cite{AB97,AB02}.

\subsection{Multipliers. The sequence of the direct method}\label{sec:2:1}

Consider a set of \emph{multipliers} $\{\Lambda_\sigma[U]\}_{\sigma=1}^N=\{\Lambda_\sigma(x, U,
\partial{U},\ldots ,\partial^l{U})\}_{\sigma=1}^N$ which, taken as factors in the linear combination of
equations of the PDE system \eqref{eq:ch1:secCL:PDEsys}, yield a \emph{divergence expression}
\begin{equation} \label{eq:ch1:secCL:CL_mixU}
\Lambda_\sigma[U] R^\sigma[U] \equiv\sg{D}_{i}\Phi^{i}[U],
\end{equation}
which holds for \emph{arbitrary} functions $U(x)$. Then on the solutions $U(x)=u(x)$ of the PDE system
\eqref{eq:ch1:secCL:PDEsys}, one has a local conservation law
\begin{equation} \label{eq:ch1:secCL:CL_mix}
\Lambda_\sigma[u] R^\sigma[u] =\sg{D}_{i}\Phi^{i}[u]=0,
\end{equation}
provided that each multiplier is non-singular. [A multiplier $\Lambda_\sigma[U]$ is called \emph{singular} if it is a singular
function when computed on solutions $U(x)=u(x)$ of a given PDE system \eqref{eq:ch1:secCL:PDEsys} (e.g., $\Lambda_\sigma[U]=
F[U]/R^\sigma[U]$). In practice, one is interested only in non-singular sets of multipliers, since singular multipliers can yield
arbitrary divergence expressions which are not conservation laws of the given PDE system. For example, taking
$\Lambda_{\sigma_1}[U]= \div \vec{\Phi}[U]/R^{\sigma_1}[U]$ and $\Lambda_{\sigma\ne\sigma_1}[U]=0$, one obtains
$\Lambda_\sigma[U] R^\sigma[U] = \div \vec{\Phi}[U]$, which does not yield a conservation law of the PDE system
\eqref{eq:ch1:secCL:PDEsys}.]

\medskip
To seek sets of multipliers $\{\Lambda_\sigma[U]\}_{\sigma=1}^N$ that yield conservation laws \eqref{eq:ch1:secCL:CL_mix} of
a given PDE system \eqref{eq:ch1:secCL:PDEsys}, one uses the following fundamental property of Euler operators
\eqref{eq:ch1:sec13:Bluman:EulerOp}.

\begin{theorem}\label{th:ch1:sec1_3_7:Euler2}
Let $x=(x^1,\ldots ,x^n)$ and $U(x)=(U^1(x),\ldots ,U^m(x))$. An expression $f[U]=f(x,U,\partial U, \ldots , \partial^l U)$,
$l\geq 0$, is annihilated by all Euler operators $\sg{E}_{U^j}$, i.e.,
\begin{equation}
\label{eq:ch1:sec13:Bluman:3} \sg{E}_{U^j} f[U]\equiv 0, \quad j = 1,\ldots,m,
\end{equation}
if and only if $f[U]$ is a divergence expression
\[
f[U]=\sg{D}_i \Phi ^i[U],
\]
for some set of fluxes $\{\Phi^i[U]\}_{i=1}^n = \{\Phi^i(x, U,
\partial{U},\ldots ,\partial^r{U})\}_{i=1}^n$.
\end{theorem}

\begin{corollary}\label{cor1}
A set of non-singular multipliers $\{\Lambda_\sigma(x,U,\partial U,\ldots ,\partial^l U)\}_{\sigma = 1}^N$ yields a local
conservation law for the PDE system  \eqref{eq:ch1:secCL:PDEsys} if and only if
\begin{equation}
\label{eq:ch1:sec13:Bluman:4} \sg{E}_{U^j} (\Lambda_\sigma(x,U,\partial U,\ldots ,\partial ^l U)R^\sigma(x,U,\partial
U,\ldots ,\partial ^kU)) = 0,\quad j = 1,\ldots ,m,
\end{equation}
hold for arbitrary functions $U(x)$.
\end{corollary}

\subsubsection*{The sequence of the direct method}
\begin{enumerate}
 \item For a given PDE system  \eqref{eq:ch1:secCL:PDEsys}, seek multipliers of the form
     $\{\Lambda_\sigma(x,U,\partial U,\ldots ,\partial ^l U)\}_{\sigma=1}^N$, for some specified order
     $l$. Choose the dependence of multipliers on their arguments so that singular multipliers do not
     arise.

 \item Solve the set of determining equations \eqref{eq:ch1:sec13:Bluman:4} for arbitrary $U(x)$ to find
     all such sets of multipliers.

 \item Find the corresponding fluxes $\Phi^{i}[U]=\Phi^{i}({x},{ U},\partial {U},\ldots , \partial ^{r} {U})$
     satisfying the identity \eqref{eq:ch1:secCL:CL_mixU}.


 \item Each set of fluxes and multipliers yields a local conservation law
\[
\sg{D}_i \Phi^{i}({x},u,\partial u,\ldots , \partial ^{r} {u}) =0,
\]
holding for all solutions $u(x)$ of the given PDE system  \eqref{eq:ch1:secCL:PDEsys}.
\end{enumerate}

In practical computations, when it is possible, it is convenient to write the given PDE system \eqref{eq:ch1:secCL:PDEsys}
in a solved form \eqref{eq:ch1:secCL:PDEsys:solved_lead} with respect to some leading derivatives. Then in Step 1 of the
direct method, one can easily avoid singular conservation law multipliers, by excluding such leading derivatives and their
differential consequences from the multiplier dependence.

\subsection{Completeness of the direct method}

The direct method outlined above can be applied to a wide class of PDE systems. Indeed, the only requirement is the
existence of all derivatives that arise in the determining equations \eqref{eq:ch1:sec13:Bluman:4}.

However, for a given PDE system, it is often important to know whether or not the direct method can yield \emph{all} its
nontrivial local conservation laws (at least up to some fixed order). Conversely, under what conditions does a set of local
multipliers yield \emph{only nontrivial} local conservation laws? We now outline the results related to these questions.

When a given PDE system is written in a solved form, one can show that the application of the direct method with multipliers
of sufficiently general dependence yields \emph{all local conservation laws} of the given PDE system of any fixed order
order $l$. In particular, the following theorem holds \cite{BCAbook}.

\begin{theorem}\label{th:ch1:secCL:PDEsys:solved_lead:CL_from_mults}
For each local conservation law $\sg{D}_{i}\Phi^i[u]=0$ of a PDE system \eqref{eq:ch1:secCL:PDEsys} written
in a solved form \eqref{eq:ch1:secCL:PDEsys:solved_lead}, there exists an equivalent local conservation law
$\sg{D}_{i}\widetilde{\Phi}^i[u]=0$ which can be expressed in a \emph{characteristic form}
\beq\label{th:ch1:secCL:PDEsys:solved_lead:CL_from_mults:char}
\sg{D}_{i}\widetilde{\Phi}^i[U] = \widetilde{\Lambda}_\sigma[U]\left(U^{j_\sigma}_{i_{\sigma,1} \ldots i_{\sigma,s}} -
G^\sigma[U]\right)
\eeq
in terms of a set of non-singular local multipliers $\{\widetilde{\Lambda}_\sigma[U]\}_{\sigma=1}^N$, with fluxes that contain no
leading derivatives $U^{j_\sigma}_{i_{\sigma,1} \ldots i_{\sigma,s}}$ nor their differential consequences.
\end{theorem}

However, even when a PDE system is written in a solved form \eqref{eq:ch1:secCL:PDEsys:solved_lead}, the correspondence
between equivalence classes of conservation laws and sets of conservation law multipliers might not be one-to-one. The
following theorem holds \cite{AB02}.

\begin{theorem}\label{th:ch1:secCL:CLsFromMults}
Suppose a PDE system \eqref{eq:ch1:secCL:PDEsys} admits a Cauchy-Kovalevskaya form
\eqref{eq:ch1:secCL:PDEsys:CKform}. Then all of its non-trivial (up to equivalence) local conservation laws
arise from multipliers. Moreover, there is a one-to-one correspondence between equivalence classes of
conservation laws and sets of conservation law multipliers $\{\Lambda_\sigma[U]\}_{\sigma=1}^N$, if the
multipliers do not involve derivatives with respect to $x^j$.
\end{theorem}

\section{Computation of fluxes of conservation laws} \label{sec:3}
\zeroall

Consider a set of non-singular multipliers $\{\Lambda_\sigma[U]=\Lambda_\sigma(x, U,
\partial{U},\ldots ,\partial^l{U})\}_{\sigma=1}^N$ that yields a divergence expression \eqref{eq:ch1:secCL:CL_mixU} for a PDE system
\eqref{eq:ch1:secCL:PDEsys} for some set of fluxes $\{\Phi^i[U]=\Phi^i(x,U,\partial U,...,\partial ^r U)\}_{i=1}^n$. Hence
one has a local conservation law \eqref{eq:ch1:secCL:CL_mix} holding on the solutions $U(x)=u(x)$ of the PDE system
\eqref{eq:ch1:secCL:PDEsys}.

Given a set of conservation law multipliers, the problem of finding the fluxes $\{\Phi^{i}[U]\}_{i=1}^n$ is
formally a problem of inversion of the divergence differential operator. Even modulo trivial conservation
laws of the first type (Definition \ref{def:2}, $H^i[u]\equiv 0$), a set of multipliers defines \emph{an
equivalence class} of conservation laws, up to free constants and arbitrary functions. For example, suppose
in a three-dimensional space $(x,y,z)$ $(n=3)$, the fluxes of a conservation law
\[
\sum_{\sigma=1}^N\Lambda_\sigma[U] R^\sigma[U] \equiv \sum_{i=1}^3 \sg{D}_{i}\Phi^{i}[U]
\]
are given by
\[
\vec{\Phi}[U]=(\Phi^{1}[U],\Phi^{2}[U],\Phi^{3}[U])={\div}^{-1}\left(\sum_{\sigma=1}^3\Lambda_\sigma[U] R^\sigma[U]\right).
\]
Then the expression
\[
\vec{\widehat{\Phi}}[U]=\vec{\Phi}[U] + \curl \vec{K},
\]
for an arbitrary smooth vector field $\vec{K}$, also yields fluxes of an equivalent conservation law. From the practical
point of view, for a given set of multipliers, it is sufficient to find any one set of corresponding fluxes.

We now consider different ways of finding fluxes of conservation laws from known multipliers.

\subsection{The direct method of flux computation}\label{subsec:meth:direct}

For conservation laws arising from simple forms of multipliers, the fluxes are most easily found by direct matching of the
two sides of equation \eqref{eq:ch1:secCL:CL_mixU} and subsequent integration by parts.

If integration by parts is not obvious, one can assume a sufficiently general dependence for the fluxes $\Phi^{i}[U]$, and look
for a particular solution of PDEs \eqref{eq:ch1:secCL:CL_mixU} for the fluxes within such an ansatz. Here it is important to note
that if the maximal order of derivatives of $U(x)$ present in the multipliers is $l$, and the maximal order in the equations
$R^\sigma[U]$ appearing in the linear combination \eqref{eq:ch1:secCL:CL_mixU} is $k$, then, without loss of generality (i.e., up
to trivial conservation laws) one may assume that the fluxes are of the form $\Phi^{i}[U]=\Phi^{i}({x},{ U},\partial {U}, \ldots
,\partial ^{r} {U})$, where $r=\max(l,k)$ \cite{olv83}. Using the independence of $\Phi^{i}[U]$ from higher derivatives, one can
subsequently split PDEs \eqref{eq:ch1:secCL:CL_mixU} into a set of simpler determining equations.

We now consider an example of a direct flux computation. Further examples are found in Section \ref{sec:6}.

\bigskip\noindent\textbf{Example.} As an example, consider the nonlinear wave equation
\begin{equation} \label{eq:ch1:secCL:FluxesHowTo:NLWave}
R[u] = u_{tt}-(c^2(u)u_x)_x =0,
\end{equation}
for arbitrary wave speed $c(u)$. For simplicity, consider multipliers of the form $\Lambda[U] = \Lambda(x,t,U)$. The determining
equations \eqref{eq:ch1:sec13:Bluman:3} yield the solution $\Lambda(x,t) = C_1 + C_2 x + C_3 t + C_4 xt$, where $C_1,\ldots,C_4$
are arbitrary constants. As a result, one obtains four linearly independent conservation laws, arising from the multipliers
$\Lambda^{(1)}=1$, $\Lambda^{(2)}=x$, $\Lambda^{(3)}=t$, $\Lambda^{(4)}=xt.$ We now determine the corresponding density-flux
pairs.

For the multiplier $\Lambda^{(1)}=1$, one obviously has
\begin{equation} \label{eq:ch1:secCL:FluxesHowTo:NLWave:Cl1f}
\Lambda^{(1)}[U]R[U] \equiv \sg{D}_t(U_{t})-\sg{D}_x(c^2(U)U_x),
\end{equation}
since PDE \eqref{eq:ch1:secCL:FluxesHowTo:NLWave} is in divergence form as it stands:
\begin{equation} \label{eq:ch1:secCL:FluxesHowTo:NLWave:Cl1}
\sg{D}_t(u_{t})-\sg{D}_x(c^2(u)u_x)=0.
\end{equation}

For the multiplier $\Lambda^{(2)}=x$, one can determine the flux and density using integration by parts:
\begin{equation} \label{eq:ch1:secCL:FluxesHowTo:NLWave:Cl2f}
\begin{array}{lcl}
\Lambda^{(2)}[U]R[U] &\equiv & x(\sg{D}_t(U_{t})-\sg{D}_x(c^2(U)U_x)) \\
&\equiv & \sg{D}_t(xU_{t}) - \sg{D}_x(x c^2(U)U_x) + c^2(U)U_x \\
&\equiv & \sg{D}_t(xU_{t}) - \sg{D}_x\left(x c^2(U)U_x - \displaystyle\int c^2(U)dU\right),
\earr
\end{equation}
and thus the corresponding conservation law is given by
\begin{equation} \label{eq:ch1:secCL:FluxesHowTo:NLWave:Cl2}
\sg{D}_t(xu_{t}) - \sg{D}_x\left(x c^2(u)u_x - \int c^2(u)du\right)=0.
\end{equation}

Similarly, for the multiplier $\Lambda^{(3)}=t$, one finds the corresponding conservation law
\begin{equation} \label{eq:ch1:secCL:FluxesHowTo:NLWave:Cl3}
\sg{D}_t(tu_{t}-u) - \sg{D}_x(t c^2(u)u_x)=0.
\end{equation}

To find the flux and density of the somewhat more complicated fourth conservation law of PDE
\eqref{eq:ch1:secCL:FluxesHowTo:NLWave} following from the multiplier $\Lambda^{(4)}=xt$, one needs to solve the flux-density
determining equation
\begin{equation} \label{eq:ch1:secCL:FluxesHowTo:NLWave:Cl4f}
\Lambda^{(4)}[U]R[U] = xt(\sg{D}_t(U_{t})-\sg{D}_x(c^2(U)U_x)) \equiv \sg{D}_t\Psi[U] + \sg{D}_x\Phi[U].
\end{equation}
Since the left-hand side of \eqref{eq:ch1:secCL:FluxesHowTo:NLWave:Cl4f} is linear in the highest derivatives $U_{tt}$ and
$U_{xx}$, one can assume that $\Psi[U]=\Psi(x,t,U,U_x,U_t)$ and $\Phi[U]=\Phi(x,t,U,U_x,U_t)$. Expanding both sides of
\eqref{eq:ch1:secCL:FluxesHowTo:NLWave:Cl4f}, one obtains
\begin{equation} \label{eq:ch1:secCL:FluxesHowTo:NLWave:Cl4f_2}
\begin{array}{lcl}
xt (U_{tt} - 2c(U)c'(U)(U_x)^2 - c^2(U)U_{xx}) &\\
\quad  =\left( \Psi_t + \Psi_U U_t + \Psi_{U_t}U_{tt} + \Psi_{U_x}U_{xt}\right) \\
\quad  + \left( \Phi_x + \Phi_U U_x + \Phi_{U_t}U_{xt} + \Phi_{U_x}U_{xx}\right).
\earr
\end{equation}
Matching the terms of the highest order derivatives $U_{tt}$, $U_{xx}$ and $U_{xt}$, one finds that
\beq \label{eq:ch1:secCL:FluxesHowTo:NLWave:Cl4f_3}
xt=\Psi_{U_t},\quad -xt c^2(U) = \Phi_{U_x}, \quad \Psi_{U_x} = - \Phi_{U_t}.
\eeq
The third equation in \eqref{eq:ch1:secCL:FluxesHowTo:NLWave:Cl4f_3} can be replaced by $\Psi_{U_x} = \Phi_{U_t}=0$, since other
choices lead to equivalent conservation laws. The first two equations in \eqref{eq:ch1:secCL:FluxesHowTo:NLWave:Cl4f_3}
yield
\beq \label{eq:ch1:secCL:FluxesHowTo:NLWave:Cl4f_4}
\Psi[U]=xtU_t + \alpha(x,t,U),\quad \Phi[U]= -xt c^2(U) U_x + \beta(x,t,U),
\eeq
for arbitrary $\alpha(x,t,U)$, $\beta(x,t,U)$. Substituting \eqref{eq:ch1:secCL:FluxesHowTo:NLWave:Cl4f_4} into the
determining equations \eqref{eq:ch1:secCL:FluxesHowTo:NLWave:Cl4f_2} and setting to zero coefficients of first-order partial
derivatives of $U$, one finds
\[
x = - \alpha_U, \quad tc^2(U)= \beta_U, \quad \alpha_t = -\beta_x.
\]
Therefore
\[
\alpha(x,t,U) = -xU + \widetilde{\alpha}(x,t), \quad \beta(x,t,U) = t\int c^2(U) dU + \widetilde{\beta}(x,t),\quad
\widetilde{\alpha}_t = -\widetilde{\beta}_x.
\]
It is evident that any choice of $\widetilde{\alpha}$ and $\widetilde{\beta}$ satisfying $\widetilde{\alpha}_t =
-\widetilde{\beta}_x$ yields an equivalent conservation law, with the simplest one having
$\widetilde{\alpha} = \widetilde{\beta} =0$. Thus, the fourth conservation law of PDE
\eqref{eq:ch1:secCL:FluxesHowTo:NLWave} is given by
\begin{equation} \label{eq:ch1:secCL:FluxesHowTo:NLWave:Cl4}
\sg{D}_t(xtu_t -xu) - \sg{D}_x\left(xt c^2(u)u_x - t\int c^2(u) du\right)=0.
\end{equation}

\subsection{The first homotopy formula}\label{subsec:meth:herem}

In the case of complicated forms of multipliers and/or equations, for the inversion of divergence operators, one can use
homotopy operators that arise in differential geometry and reduce the problem of finding fluxes to a problem of integration
in single-variable calculus. In this section, we present the first set of such formulas, following \cite{hereman1}.

\begin{definition} The \emph{$n$-dimensional higher Euler operator} with respect to a function $U(x^1,...,x^n)$ is given by
\beq \label{eq:ch1:secCL:HigherEul_nD}
\sg{E}^{(s_1,\ldots , s_n)}_{U} = \sum_{k_1=s_1}^\infty \ldots \sum_{k_n=s_n}^\infty \left(\barr{k_1}\\{s_1}\earr\right) \ldots
\left(\barr{k_n}\\{s_n}\earr\right) \; \sg{D}_{1}^{k_1-s_1}\ldots\sg{D}_{n}^{k_n-s_n} \frac{\partial }{\partial
U^{(k_1+\ldots+k_n)} },
\eeq
\end{definition}
where one denotes
\[
U^{(k_1+\ldots+k_n)}=\dfrac{\partial^{k_1+\ldots+k_n} U}{\partial^{k_1} x^1\ldots \partial^{k_n} x^n}.
\]
Note that $\sg{E}^{(0, 0)}_{U}=\sg{E}_U $ is the usual Euler operator \eqref{eq:ch1:sec13:Bluman:EulerOp}
with respect to $U$.

\begin{definition} Let $x = (x^1,\ldots ,x^n)$ and $U=(U^1(x),\ldots, U^m(x))$. For an expression $f[U]=f(x,U,\partial U,\ldots)$, the
\emph{$n$-dimensional homotopy operator} is defined through its $n$ components corresponding to $n$ respective independent
variables $x^i$, $i=1,\ldots,n$, as follows:
\beq \label{eq:ch1:secCL:Homotopy_2D_xI}
\sg{H}^{(x^i)}(f[U]) = \int_0^1 \sum_{j=1}^m \sg{I}_j^{(x^i)}(f[\widehat{U}])\Big|_{\widehat{U}=\lambda
U}\frac{d\lambda}{\lambda},
\eeq
with
\beq \label{eq:ch1:secCL:Homotopy_2D_It}
\barr
\sg{I}_{j}^{(x^i)}(f[\widehat{U}])  =& \displaystyle \sum_{s_1=0}^\infty \ldots \sum_{s_n=0}^\infty \left(
\dfrac{1+s_i}{1+s_1+\ldots+s_n}\right) \\[3ex]
&\displaystyle \times \sg{D}_1^{s_1}\ldots \sg{D}_n^{s_n}\left( \displaystyle \widehat{U}^j \sg{E}^{(s_1,\ldots,s_{i-1},
\,1+ s_i,\,s_{i+1},\ldots, s_n)}_{\widehat{U}^j}(f[\widehat{U}])\right),
\earr
\eeq
$j=1,\ldots, m$.
\end{definition}

The following theorem holds \cite{hereman1}.

\begin{theorem}\label{th:ch1:secCL:Homotopy1}
Suppose $f[U]$ is a divergence expression
\[
f[U] = \div \vec{\Phi} [U] = \sg{D}_1 \Phi^1[U] + \ldots + \sg{D}_n \Phi^n[U],
\]
and $f[0]=0$. Then the fluxes $\Phi^i[U]$ are given by
\beq \label{eq:ch1:secCL:th:Homotopy1}
\Phi^i[U]=\sg{H}^{(x^i)}(f[U]),\quad i=1,\ldots,n.
\eeq
up to corresponding fluxes of a trivial conservation law, provided that the integrals
\eqref{eq:ch1:secCL:Homotopy_2D_xI} converge.
\end{theorem}

Two examples of the use of the first homotopy formula are now considered. Further examples are found in Section \ref{sec:6}.

\bigskip\noindent\textbf{Example 1.} As a first example, consider the Korteweg-de Vries (KdV) equation
\beq\label{eq:ch1:sec13:KdV:secCL}
u_t+uu_x +u_{xxx}=0.
\eeq
One can show that PDE \eqref{eq:ch1:sec13:KdV:secCL} has a local conservation law arising from the multiplier
$\Lambda[U]=U$, i.e.,
\[
f[U]=U(U_t+UU_x +U_{xxx}) \equiv \sg{D}_t \Psi[U] + \sg{D}_x \Phi[U].
\]
is a divergence expression. We now reconstruct the corresponding density $\Psi[U]$ and flux $\Phi[U]$. First compute
\[
I^{(t)}(f[\widehat{U}]) = \sg{D}^0_t \sg{D}^0_x \left(\widehat{U} \sg{E}^{(1,
0)}_{\widehat{U}}(f[\widehat{U}]) \right) = \widehat{U}^2,
\]
\[
\begin{array}{lcl}
I^{(x)}(f[\widehat{U}]) &=& \sg{D}^0_t \sg{D}^2_x \left(\widehat{U} \sg{E}^{(0, 3)}_{\widehat{U}}(f[\widehat{U}]) \right) +
\sg{D}^0_t \sg{D}^1_x \left(\widehat{U}
\sg{E}^{(0, 2)}_{\widehat{U}}(f[\widehat{U}]) \right) + \sg{D}^0_t \sg{D}^0_x \left(\widehat{U} \sg{E}^{(0, 1)}_{\widehat{U}}(f[\widehat{U}]) \right) \\[1.5ex]
&=& \widehat{U}^3-\widehat{U}^2_x + 2\widehat{U}\widehat{U}_{xx}.
\earr
\]
Consequently,
\[
\sg{H}^{(t)}(f[U]) = \int_0^1 (\lambda U)^2 \frac{d\lambda}{\lambda} = \tfrac{1}{2}U^2,
\]
\[
\sg{H}^{(x)}(f[U]) = \int_0^1 \left(\lambda^3U^3-\lambda^2U_x^2 + 2\lambda^2 UU_{xx} \right) \frac{d\lambda}{\lambda} =
\tfrac{1}{3}U^3-\tfrac{1}{2}U_x^2 + UU_{xx},
\]
The corresponding conservation law is given by
\beq\label{eq:herem:cl:KdVU}
\sg{D}_t \left(\tfrac{1}{2}u^2\right) + \sg{D}_x \left(\tfrac{1}{3}u^3-\tfrac{1}{2}u_x^2 + uu_{xx}\right)=0.
\eeq

\bigskip\noindent\textbf{Example 2.} Consider the 2-dimensional $G$-equation
\beq\label{eq:ch1:Symm_CL:ancoway:eg_G}
R[g] = g_t- |\grad g| \equiv g_t-\sqrt{g_{x}^2+g_y^2}=0,
\eeq
which describes flame propagation in a static gas (see, e.g., \cite{OberlackWenzelPeters}). Here $g(t,x,y)=0$ implicitly
defines the position of the flame surface at time $t$, and the surface advances at a constant speed in the normal direction.
Let $(x^1,x^2)=(x,y)$. In particular, one can show that the PDE \eqref{eq:ch1:Symm_CL:ancoway:eg_G} has the following
multipliers of conservation laws:
\beq\label{eq:ch1:Symm_CL:ancoway:eg_G:Muls}
\Lambda^{(1)}[G]=\frac{1}{G^3_y}(G_x G_{yy}-G_yG_{xy}),\quad \Lambda^{(2)}[G]= H(G_x , G_y) (G_{xx}G_{yy}-G_{xy}^2),
\eeq
where $H(G_x , G_y)$ is an arbitrary function. For more details, see \cite{ChevOberl}.

First consider the multiplier $\Lambda^{(1)}[G]$ in \eqref{eq:ch1:Symm_CL:ancoway:eg_G:Muls}. From
\eqref{eq:ch1:secCL:Homotopy_2D_It}, one obtains
\[
\sg{I}^{(t)}(\Lambda^{(1)}[G]R[G]) = \frac{G}{G_x^3}(G_xG_{xy}-G_yG_{xx}),
\]
which is an invariant expression under the transformation $\widehat{U}=\lambda U$. The same is true about expressions
$\sg{I}^{(x)}(\Lambda^{(1)}[G]R[G])$ and $\sg{I}^{(y)}(\Lambda^{(1)}[G]R[G])$. Hence the integral homotopy operators
\eqref{eq:ch1:secCL:Homotopy_2D_xI} yield a divergent integral $\int_0^1 d\lambda/\lambda$.

Second, consider the multiplier $\Lambda^{(2)}[G]$ in \eqref{eq:ch1:Symm_CL:ancoway:eg_G:Muls}. Since the corresponding
divergence expression
\[
f[G]=\Lambda^{(2)}[G]R[G] = H(G_x , G_y) (G_t-\sqrt{G_{x}^2+G_y^2}) (G_{xx}G_{yy}-G_{xy}^2)
\]
contains an arbitrary function $H(G_x , G_y)$, the corresponding integrals
\eqref{eq:ch1:secCL:Homotopy_2D_xI} cannot yield closed-form expressions.

It follows that the present flux construction method cannot be used for either of the multipliers
\eqref{eq:ch1:Symm_CL:ancoway:eg_G:Muls} of the $G$-equation \eqref{eq:ch1:Symm_CL:ancoway:eg_G}.

\subsection{The second homotopy formula}\label{subsec:meth:BA}

An alternative set of integral formulas for fluxes of local conservation laws was developed by Bluman and Anco \cite{AB02}.
Here we present these formulas in a different form, which is symmetric with respect to independent variables, and also is
simpler for practical computations and software implementation.

It takes a somewhat bigger effort to implement the current method in symbolic computation software than to implement the
method presented in Section \ref{subsec:meth:herem}. However the formulas given in the current section have two important
practical advantages over the ones presented in Section \ref{subsec:meth:herem}.
\begin{enumerate}
    \item[(i)] They can be applied to divergence expressions $f[U]$ with $f[0]\ne0$.

    \item[(ii)] They have flexibility that can be used to always avoid divergent integrals.
\end{enumerate}

The following theorem holds.

\begin{theorem}\label{th:ch1:secCL:HomotopyACBA}
Suppose a PDE system \eqref{eq:ch1:secCL:PDEsys} has a local conservation law arising from a set of multipliers
$\{\Lambda_\sigma[U]\}_{\sigma=1}^N$, i.e., \eqref{eq:ch1:secCL:CL_mix} holds. Then the fluxes $\{\Phi^i[{U}]\}_{i=1}^N$ of
that conservation law are given by
\beq\label{eq:ch1:secCL:FluxesHowTo:homot:flux_int}
\barr
\Phi^i[U]=&\Phi^i[\widetilde{U}]+\displaystyle \int_0^1\left(S^i\left[U-\widetilde{U},\Lambda[\lambda U +
(1-\lambda)\widetilde{U}]; R[\lambda U +(1-\lambda)\widetilde{U}]\right]\right.\\[2ex]
&\left.+\widetilde{S}^i\left[U-\widetilde{U},R[\lambda U +(1-\lambda)\widetilde{U}]; \Lambda[\lambda U
+(1-\lambda)\widetilde{U}]\right]\right)d\lambda,\quad i=1,\ldots,n,
\earr
\eeq
up to fluxes of a trivial conservation law. In \eqref{eq:ch1:secCL:FluxesHowTo:homot:flux_int}, operators
$S^i$, $\widetilde{S}^i$ are defined by their action on arbitrary functions $V=(V^1(x),\ldots,V^m(x))$,
$W=(W_1(x),\ldots,W_N(x))$, $\widetilde{V}=(\widetilde{V}^1(x),\ldots,\widetilde{V}^m(x))$ and
$\widetilde{W}=(\widetilde{W}_1(x),\ldots,\widetilde{W}_N(x))$ as follows:
\beq\label{eq:ch1:secCL:FluxesHowTo:homot:WLV_S}
S^i[V,W; R[U]] = \sum_{p=0}^{k-1} \sum_{q=0}^{k-p-1}(-1)^q \left( \sg{D}_{i_1} \ldots \sg{D}_{i_p}
V^\rho\right) \sg{D}_{j_1} \ldots \sg{D}_{j_q}\left(W_\sigma \frac{\partial R^\sigma[U]}{\partial U^\rho_{j_1
\ldots j_q i i_1 \ldots i_p}}\right),
\eeq
\beq\label{eq:ch1:secCL:FluxesHowTo:homot:WLV_Stilda}
\widetilde{S}^i[\widetilde{V},\widetilde{W}; \Lambda[U]] = \sum_{p=0}^{l-1} \sum_{q=0}^{l-p-1}(-1)^q \left( \sg{D}_{i_1} \ldots
\sg{D}_{i_p} \widetilde{V}^\rho\right) \sg{D}_{j_1} \ldots \sg{D}_{j_q}\left(\widetilde{W}^\sigma \frac{\partial
\Lambda_\sigma[U]}{\partial U^\rho_{j_1 \ldots j_q i i_1 \ldots i_p}}\right),
\eeq
where $k$ is the order of the given PDE system $\PDEs{R}{x}{u}$ \eqref{eq:ch1:secCL:PDEsys}, $l$ is the maximal order of the
derivatives appearing in the multipliers $\{\Lambda_\sigma[U]\}_{\sigma=1}^N$, and $j_1 \ldots j_q$, $i_1 \ldots i_p$ are ordered
combinations of indices such that $1\leq j_1\leq  \ldots \leq j_q\leq i\leq i_1\leq  \ldots \leq i_p\leq n$.

In \eqref{eq:ch1:secCL:FluxesHowTo:homot:flux_int}, $\widetilde{U}=\widetilde{U}(x)$ is an arbitrary fixed function, and
$\Phi^i[\widetilde{U}]$ are quantities satisfying
\beq\label{th:ch1:secCL:HomotopyACBA:tilda}
\sg{D}_i\Phi^i[\widetilde{U}]=\Lambda_\sigma[\widetilde{U}]R^\sigma[\widetilde{U}]=:F(x).
\eeq
\end{theorem}
The proof of Theorem \ref{th:ch1:secCL:HomotopyACBA} is presented in Appendix \ref{app:a}.

\medskip
In  \eqref{eq:ch1:secCL:FluxesHowTo:homot:flux_int}, $\widetilde{U}$ is an arbitrary function of $x$, chosen so that the
integral \eqref{eq:ch1:secCL:FluxesHowTo:homot:flux_int} converges. Different choices of $\widetilde{U}$ yield fluxes of
equivalent conservation laws. One normally chooses $\widetilde{U}=0$ (provided that the integral converges). Once
$\widetilde{U}=\widetilde{U}(x)$ has been chosen, the functions $\{\Phi^i[\widetilde{U}]\}_{i=1}^N$ can be found by direct
integration from the divergence relation \eqref{th:ch1:secCL:HomotopyACBA:tilda} by integrating $F(x)$. For example, one can
choose
\[
\Phi^1[\widetilde{U}]=\int F(x) dx^1,\quad \Phi^2[\widetilde{U}]=\ldots=\Phi^n[\widetilde{U}]=0.
\]

It is important that unlike the case with the first homotopy formula (Section \ref{subsec:meth:herem}), the proof of Theorem
\ref{th:ch1:secCL:HomotopyACBA} does not require that the divergence expression $\Lambda_\sigma[U] R^\sigma[U]$ vanishes for
$U=0$.

The formula \eqref{eq:ch1:secCL:FluxesHowTo:homot:flux_int} is useful in many practical situations when the
integrals can be explicitly evaluated.

Two examples of the use of the second homotopy formula are now considered; further examples are found in Section \ref{sec:6}.

\bigskip\noindent\textbf{Example 1.} As a first example, consider again the KdV equation \eqref{eq:ch1:sec13:KdV:secCL} with multiplier
$\Lambda[U]=U$. Using $\widetilde{U}=0$, one finds $\Phi^1[0]=\Phi^2[0]=0$, and hence the formulas
\eqref{eq:ch1:secCL:FluxesHowTo:homot:flux_int} yield exactly the same fluxes as the first homotopy method (Section
\ref{subsec:meth:herem}), i.e., the conservation law \eqref{eq:herem:cl:KdVU}.

Alternatively, using $\widetilde{U}=x$, one finds $\Phi^1[\widetilde{U}]=tx^2, \Phi^2[\widetilde{U}]=0$, and hence
\[
\barr
\Phi^1[U] = tx^2+\tfrac{1}{2}(U^2-x^2),\\[2ex]
\Phi^2[U] = \tfrac{1}{2} - \tfrac{1}{3}x^3 + \tfrac{1}{3}U^3 - \tfrac{1}{2}U_x^2 + UU_{xx},
\earr
\]
which are a density and a flux of a conservation law equivalent to \eqref{eq:herem:cl:KdVU}.

\bigskip\noindent\textbf{Example 2.} Consider the 2-dimensional $G$-equation
\eqref{eq:ch1:Symm_CL:ancoway:eg_G} and its local conservation law multipliers \eqref{eq:ch1:Symm_CL:ancoway:eg_G:Muls}. For
the multiplier $\Lambda^{(1)}[G]$, it again yields a divergent integral when $\widetilde{U}=0$. The singularity can be
removed by taking e.g. $\widetilde{U}=x$, however even this simple choice leads to an integral with a highly complicated
integrand. For the multiplier $\Lambda^{(2)}[G]$ involving an arbitrary function, formulas
\eqref{eq:ch1:secCL:FluxesHowTo:homot:flux_int} do not yield a closed-form expression.

\subsection{Computation of fluxes of conservation laws of scaling-invariant PDE systems}\label{subsec:meth:Anco}

Consider a PDE system that has a scaling symmetry
\beq\label{eq:ch1:Symm_CL:sym_adj_CL:scal_sym}
\sg{X}_s[u] = p^{(i)} x^i\frac{\partial}{\partial x^i} + q^{(\rho)} u^\rho\frac{\partial}{\partial u^\rho},
\eeq
where $p^{(i)}$ and $q^{(\rho)}$ are constant scaling weights of independent and dependent variables, respectively, i.e., the
generator \eqref{eq:ch1:Symm_CL:sym_adj_CL:scal_sym} corresponds to the global transformation group
\beq\nonumber
\barr
x^i\rightarrow \tilde{x}^i=e^{\varepsilon \sg{X}_s[u]}x^i = e^{\varepsilon p^{(i)}}x^i,\\[2ex]
u^\rho\rightarrow \tilde{u}^\rho=e^{\varepsilon \sg{X}_s[u]}u^\rho = e^{\varepsilon q^{(\rho)}}u^\rho.
\earr
\eeq
The scaling symmetry generator \eqref{eq:ch1:Symm_CL:sym_adj_CL:scal_sym} can be written in evolutionary form as
\beq\label{eq:ch1:Symm_CL:sym_adj_CL:scal_sym:char}
\sg{\hat{X}}_s[u] = \hat{\eta}[u]\frac{\partial}{\partial u^\rho} =\left(q^{(\rho)} u^\rho - p^{(i)} x^i
u^\rho_i\right)\frac{\partial}{\partial u^\rho}.
\eeq
Assuming that the given PDE system can be written in a solved form \eqref{eq:ch1:secCL:PDEsys:solved_lead}, we first note that
the scaling homogeneity of each PDE of the given system follows from considering the action of the scaling symmetry
\eqref{eq:ch1:Symm_CL:sym_adj_CL:scal_sym} on the leading derivative in each $R^\sigma [u]$. In particular, one has
\[
\sg{X}_s ^{(k)}[U] R^\sigma [U] = r^{(\sigma)} R^\sigma[U],
\]
where $r^{(\sigma)}=\const$ is the scaling weight of each PDE $R^\sigma [U]$.

Now suppose the given PDE system \eqref{eq:ch1:secCL:PDEsys:solved_lead} has a conservation law
\beq\label{eq:CL:anco}
\Lambda_\sigma[u] R^\sigma[u] =\sg{D}_{i}\Phi^{i}[u]=0,
\eeq
with fluxes $\Phi^i[U]= \Phi^i(x,U,\partial U, \ldots , \partial^l U)$. Suppose that this conservation law is scaling-invariant,
and moreover, homogeneous under the scaling symmetry \eqref{eq:ch1:Symm_CL:sym_adj_CL:scal_sym}, i.e.,
\beq\label{eq:scal_sym:CLweight}
\sg{X}_s^{(l)}[U]\sg{D}_i\Phi^i[U] = P \sg{D}_i\Phi^i[U],
\eeq
where $\sg{X}_s^{(l)}$ is a corresponding prolongation of the scaling symmetry generator
\eqref{eq:ch1:Symm_CL:sym_adj_CL:scal_sym}, and $P=\const$ is a scaling weight of the conservation law.

Then one can show \cite{BCAbook} that each of the multipliers $\Lambda_\sigma [U]$ appearing in \eqref{eq:CL:anco} is homogeneous
under the scaling symmetry \eqref{eq:ch1:Symm_CL:sym_adj_CL:scal_sym}. In particular, $\sg{X}_s^{(l)}[U]\Lambda_\sigma [U] =
s_{(\sigma)} \Lambda_\sigma[U]$, where $s_{(\sigma)}=P - r^{(\sigma)}$ is the scaling weight of each $\Lambda_\sigma[U]\ne 0$.

\begin{definition}\label{def:noncr}
A conservation law \eqref{eq:CL:anco} invariant with respect to a scaling symmetry $\sg{X}_s[u]$
\eqref{eq:ch1:Symm_CL:sym_adj_CL:scal_sym} is called \emph{noncritical with respect to $\sg{X}_s[u]$} if
\beq\label{eq:ch1:Symm_CL:sym_adj_CL:noncrit_CL}
\chi=s_{(\sigma)}+r^{(\sigma)} + \sum_{i=1}^{n} p^{(i)}\neq 0
\eeq
for each $\sigma$ such that $\Lambda_\sigma [U]\neq 0$.
\end{definition}

\subsubsection*{The formula for fluxes}

It follows that fluxes of noncritical homogeneously scaling conservation laws of scaling-invariant PDE systems can be found
through a simple formula that involves no integration! The following theorem holds \cite{AncoScal,BCAbook}.

\begin{theorem} \label{th:ch1:Symm_CL:sym_adj_Anco}
Suppose the PDE system in the solved form \eqref{eq:ch1:secCL:PDEsys:solved_lead} has a scaling symmetry
\eqref{eq:ch1:Symm_CL:sym_adj_CL:scal_sym:char}, and a conservation law \eqref{eq:ch1:secCL:def_CL_nvar}, which is homogeneous
under the scaling symmetry \eqref{eq:ch1:Symm_CL:sym_adj_CL:scal_sym}, and is noncritical with respect to it. Then the fluxes
$\Phi^i[U]$ of such conservation law are given by
\beq\label{eq:ch1:sec13:Bluman136:WLV_phi}
\Phi^i[U] = \sum_{p=0}^{k-1} \sum_{q=0}^{k-p-1}(-1)^q \left( \sg{D}_{i_1} \ldots \sg{D}_{i_p}
\hat{\eta}^\rho\right) \sg{D}_{j_1} \ldots \sg{D}_{j_q}\left(\Lambda_\sigma [U] \frac{\partial
R^\sigma[U]}{\partial U^\rho_{j_1 \ldots j_q i i_1 \ldots i_p}}\right),
\eeq
up to fluxes of a trivial conservation law. In \eqref{eq:ch1:sec13:Bluman136:WLV_phi}, $k$ is the maximal
order of derivatives appearing in the PDE system \eqref{eq:ch1:secCL:PDEsys}, $l$ is the maximal order of
derivatives appearing in the multipliers $\{\Lambda_\sigma[U]\}_{\sigma=1}^N$, and $j_1 \ldots j_q$ and $i_1
\ldots i_p$ are ordered combinations of indices such that $1\leq j_1\leq \ldots \leq j_q\leq  i\leq i_1\leq
\ldots \leq i_p\leq n$.
\end{theorem}

\begin{remark}
From the proof of Theorem \ref{th:ch1:Symm_CL:sym_adj_Anco} it follows that the formulas \eqref{eq:ch1:sec13:Bluman136:WLV_phi}
does not yield fluxes of critical conservation laws. In particular, for critical conservation laws, formulas
\eqref{eq:ch1:sec13:Bluman136:WLV_phi} yield fluxes of trivial conservation laws.

In practice, if a PDE system has two or more scaling symmetries, it can be the case that a particular conservation law is
critical with respect to one scaling and noncritical with respect to the other, as is the case in the Example 2 below.
\end{remark}

\begin{remark}
For first- and second-order PDE systems, the flux formula \eqref{eq:ch1:sec13:Bluman136:WLV_phi} takes a particularly simple
form \cite{AncoScal}. For second-order PDE systems, it becomes
\beq \label{eq:ch1:Symm_CL:AncoScal:order2}
{\Phi}^i[U] = \hat{\eta}^\rho[U]\Lambda_\sigma[U] \frac{\partial R^\sigma[U]}{\partial u^\rho_{i}} + (\sg{D}_s\hat{\eta}^\rho[U])
\Lambda_\sigma[U] \frac{\partial R^\sigma[U]}{\partial u^\rho_{is}} - \hat{\eta}^\rho[U] \sg{D}_j\left(\Lambda_\sigma[U]
\frac{\partial R^\sigma[U]}{\partial u^\rho_{ji}}\right),
\eeq
where the summation is taken over repeated indices satisfying $s\geq i$ and $j\leq i$. For first-order PDE
systems, it further reduces to
\beq \label{eq:ch1:Symm_CL:AncoScal:order1}
{\Phi}^i[U] = \hat{\eta}^\rho[U]\Lambda_\sigma[U] \frac{\partial R^\sigma[U]}{\partial u^\rho_{i}}.
\eeq
\end{remark}

\begin{remark}
In general, one can obtain fluxes of conservation laws using formula \eqref{eq:ch1:sec13:Bluman136:WLV_phi} not only for scaling
symmetries, but for any pair consisting of a local symmetry $\sg{\hat{X}}_s[u] = \hat{\eta}[u]\frac{\partial}{\partial u^\rho}$
(in the evolutionary form) and a set of local conservation law multipliers $\{\Lambda_\sigma[U]\}_{\sigma=1}^N$. Details are
given in Appendix \ref{subsec:meth:PAIR} (see also \cite{AB97,BCAbook}).
\end{remark}

\bigskip\noindent\textbf{Example 1.} As a first example, we consider again the KdV equation \eqref{eq:ch1:sec13:KdV:secCL} with
multiplier $\Lambda[U]=U$. The PDE \eqref{eq:ch1:sec13:KdV:secCL}  has a scaling symmetry
\beq\label{eq:scal_sym:KdV}
\sg{X}_s[u] = x\frac{\partial}{\partial x}+3t\frac{\partial}{\partial t} -2u \frac{\partial}{\partial u}.
\eeq
Using the formula \eqref{eq:ch1:sec13:Bluman136:WLV_phi} with $(x^1,x^2)=(t,x)$, one obtains the density and
the flux
\[
\barr
\Phi^1[U] =& (2U+3tU_t+xU_x)U,\\
\Phi^2[U] =& U(3tU_{txx} + xU_{xxx})-3tU_xU_{tx} \\
 &+3 (2U+tU_t)U_xx +   U^2( 2U+3tU_t)+U_x(xU^2-3U_x),
\earr
\]
i.e., a conservation law
\beq\label{eq:scal_sym:KdV:CL}
\barr
\sg{D}_t\Big((2u+3tu_t+xu_x)u\Big) + \sg{D}_x\Big(u(3tu_{txx} + xu_{xxx})-3tu_xu_{tx} \\
 \qquad +3 (2u+tu_t)u_xx +   u^2( 2u+3tu_t)+u_x(xu^2-3u_x)\Big)=0,
\earr
\eeq
equivalent to the conservation law \eqref{eq:herem:cl:KdVU}.

\bigskip\noindent\textbf{Example 2.} Consider the 2-dimensional $G$-equation
\eqref{eq:ch1:Symm_CL:ancoway:eg_G} and its local conservation law multipliers \eqref{eq:ch1:Symm_CL:ancoway:eg_G:Muls}. The
$G$-equation \eqref{eq:ch1:Symm_CL:ancoway:eg_G} is a first-order PDE with obvious scaling symmetries
\beq\label{eq:scal_sym:G}
\sg{X}_1 = g\frac{\partial}{\partial g}, \quad \sg{X}_2=t\frac{\partial}{\partial t} + x\frac{\partial}{\partial
x}+y\frac{\partial}{\partial y}.
\eeq
Denoting $(x^1,x^2,x^3)\equiv(t,x,y)$, for $\sg{X}_1$, one has $q=1$, $p^{(1)}=p^{(2)}=p^{(3)}=0$; for $\sg{X}_2$, $q=0$,
$p^{(1)}=p^{(2)}=p^{(3)}=1$.

Consider the conservation of the $G$-equation \eqref{eq:ch1:Symm_CL:ancoway:eg_G} law with multiplier
$\Lambda^{(1)}[G]=\frac{1}{G^3_y}(G_x G_{yy}-G_yG_{xy})$ \eqref{eq:ch1:Symm_CL:ancoway:eg_G:Muls}. The function
$\Lambda^{(1)}[G]$ is evidently homogeneous under both symmetries $\sg{X}_1$ and $\sg{X}_2$ \eqref{eq:scal_sym:G}, with
respective scaling weights $s_1=-1$ and $s_2=0$. The corresponding scaling weights of the $G$-equation
\eqref{eq:ch1:Symm_CL:ancoway:eg_G} are $r_1=1$ and $r_2=-1$.

Using Definition \ref{def:noncr}, one finds $\chi_1=-1+1=0$, and $\chi_2=0-1+3=2$, hence the conservation law with multiplier
$\Lambda^{(1)}[G]$ is critical with respect to the scaling symmetry $\sg{X}_1$ and noncritical with respect to $\sg{X}_2$.
Indeed, one can directly check that the conservation law obtained by formulas \eqref{eq:ch1:sec13:Bluman136:WLV_phi} using the
symmetry $\sg{X}_1$,
\[
\sg{D}_t \left(g\Lambda^{(1)}[g]\right) - \sg{D}_x \left(\frac{gg_{x}\Lambda^{(1)}[g]}{\sqrt{g_{x}^2+g_{y}^2}}\right)
-\sg{D}_y \left(\frac{gg_{y}\Lambda^{(1)}[g]}{\sqrt{g_{x}^2+g_{y}^2}}\right) =0,
\]
is trivial. The correct conservation law fluxes corresponding to the multiplier $\Lambda^{(1)}[G]$ can be found using the
symmetry $\sg{X}_2$; such conservation law is given by
\beq
\barr
\sg{D}_t\left(\hat{\eta}[g]\Lambda^{(1)}[g]\right) - \sg{D}_x\left(\dfrac{\hat{\eta}[g]g_x\Lambda^{(1)}[g]}{\sqrt{g_x^2+g_y^2}}\right) -\sg{D}_y\left(\dfrac{\hat{\eta}[g]g_y\Lambda^{(1)}[g]}{\sqrt{g_x^2+g_y^2}}\right) =0,\\[2ex]
\hat{\eta}[g]=-tg_t-xg_x-yg_y.
\earr
\eeq

\section{Comparison of the four flux computation methods} \label{sec:5}
\zeroall

In Table \ref{table:comp}, we briefly outline the characteristics of the four methods of flux computation discussed above.

\begin{center}\renewcommand{\arraystretch}{1.1}\refstepcounter{table}\label{table:comp}
Table~\thetable: Comparison of Four Methods of Flux Computation of Section \ref{sec:3}.
\\[2ex]\small
\begin{tabular}{|l|l|l|}
\hline
\hfill Method$ \hfill$&\hfill Applicability $ \hfill$&\hfill Computational complexity$ \hfill$\\

\hline \hline

\mbox{\parbox[t]{2cm}{Direct\\(Section \ref{subsec:meth:direct})}} & \mbox{\parbox[t]{8cm}{Simpler multipliers/PDE systems, which may involve arbitrary functions.}} &  \mbox{\parbox[t]{4cm}{Solution of an overdetermined linear PDE system for
fluxes.}}\\[7ex]

\hline

\hline \mbox{\parbox[t]{2cm}{Homotopy 1\\(Section \ref{subsec:meth:herem})}} & \mbox{\parbox[t]{8cm}{ Complicated
multipliers/PDEs, not involving arbitrary
functions.\\
The divergence expression must vanish for $U=0$.\\
For some conservation laws, this method can yield divergent integrals.}} & \mbox{\parbox[t]{4cm}{One-dimensional integration.}}\\

\hline \mbox{\parbox[t]{2cm}{Homotopy 2\\(Section \ref{subsec:meth:BA})}} & \mbox{\parbox[t]{8cm}{ Complicated
multipliers/PDEs, not involving arbitrary
functions.}}& \mbox{\parbox[t]{4cm}{One-dimensional integration.}}\\

\hline \mbox{\parbox[t]{2cm}{Scaling symmetry\\(Section \ref{subsec:meth:Anco})}} & \mbox{\parbox[t]{8cm}{
Complicated multipliers/PDEs, may involve arbitrary
functions.\\
Scaling-homogeneous PDEs and multipliers. \\
Noncritical conservation laws. }}& \mbox{\parbox[t]{4cm}{Repeated differentiation.}}\\[10ex] \hline
\end{tabular}
\end{center}

One can see that there is no ``preferred" method of finding fluxes that is simple to use and sufficiently
general to be recommended for a generic conservation law of a nonlinear PDE system. The following
observations and recommendations are based on the author's experience.

\begin{itemize}
  \item If a given PDE system and multipliers do not involve arbitrary functions, one usually uses one of
      the two integral methods. Though the second integral method is more general, the first one is also
      sometimes used because it can yield simpler flux expressions.
  \item If a given PDE system and/or multipliers are not very complicated but involve arbitrary
      constitutive functions, one normally attempts the direct method of flux computation.
  \item When a given PDE system is scaling-invariant and the conservation law is scaling-homogeneous, then the scaling
      symmetry method is the method of choice, since it involves simplest computations. (However, as seen in the KdV example
      \eqref{eq:scal_sym:KdV:CL}, when some $p^{(i)}\ne0$, the densities and fluxes computed through the fourth method can
      contain multiple unnecessary terms corresponding to trivial conservation laws.)
  \item Another instance when one needs to use the fourth method is the case of complicated
      scaling-homogeneous PDEs and/or multipliers involving arbitrary functions. In this situation, the
      scaling symmetry method is the only systematic method available.
\end{itemize}

In the following Section \ref{sec:5}, we consider a software implementation of the four above-discussed flux computation
methods, applied to several practical examples.

\section{Symbolic software implementation of flux computation methods. Examples} \label{sec:6}
\zeroall

The direct method of construction of conservation laws (Section \ref{sec:2:1}) is algorithmic and has been
implemented in symbolic software\footnote{In particular, in the software by T. Wolf \cite{t_wolf2} written
for REDUCE computer algebra system.}. Here we discuss its \verb|Maple| implementation in the symbolic package
\verb|GeM| \cite{GeM} written by the author. The package \verb|GeM| also contains routines for the four
methods of flux reconstruction considered in this paper.

\subsection{General conservation laws of the nonlinear wave equation}

Here we seek local conservation laws for the nonlinear wave equation \eqref{eq:ch1:secCL:FluxesHowTo:NLWave} arising from
the multipliers of the form $\Lambda[U] = \Lambda(x,t,U)$ for an arbitrary $c(u)$.

One defines the variables and the PDE \eqref{eq:ch1:secCL:FluxesHowTo:NLWave} in \verb|GeM| using the following \verb|Maple|
commands:
\beq\label{eq:wave:maplecomm1}
\barr
\verb"with(GeM):"\\
\verb"gem_decl_vars(indeps=[x,t], deps=[U(x,t)], freefunc=[C(U(x,t)));"\\
\verb"gem_decl_eqs([diff(U(x,t),t,t)=diff(C(U(x,t))^2*diff(U(x,t),x) ,x)],"  \\
\qquad \verb"solve_for=[diff(U(x,t),t,t)]);"
\earr
\eeq
The option \verb"solve_for" is used to specify a set of leading derivatives the given PDE systems can be solved for. Note that
the direct method of conservation law computation \emph{does not} require the equations to be written in the solved form, and
therefore the specification of the option \verb"solve_for" is not necessary. However in \eqref{eq:wave:maplecomm1}, this option
\verb"solve_for" was specified, since it will be used later in the flux computation routine, which automatically verifies the
correctness of flux computations, which is accomplished by explicitly checking that the conservation law divergence expression
vanishes on solutions $U(x)=u(x)$ of the given PDE \eqref{eq:ch1:secCL:FluxesHowTo:NLWave}.

The set of determining equations for the local conservation law multipliers is obtained and simplified using the routines
\beq \label{eq:wave:maplecomm2}
\barr
\verb"det_eqs:=gem_conslaw_det_eqs([x,t, U(x,t)]):"\\
\verb"CL_multipliers:=gem_conslaw_multipliers();"\\
\verb"simplified_eqs:=DEtools[rifsimp](det_eqs,"\\
\qquad  \verb"CL_multipliers, mindim=1, arbitrary=[C(U)]);"
\earr
\eeq
The first command in \eqref{eq:wave:maplecomm2} sets up the set of local conservation law multipliers determining equations
\eqref{eq:ch1:sec13:Bluman:4}, and splits them. The splitting is done using the fact that the determining equations
\eqref{eq:ch1:sec13:Bluman:4} are polynomial expressions in terms of derivatives $U$ of orders four and higher, and such
derivatives are linearly independent. This yields an overdetermined system of 7 determining equations for the unknown local
multiplier $\Lambda[U]$. After using the \verb"rif" reduction algorithm, the system reduces to three PDEs
\[
\dfrac{\partial^2 \Lambda[U]}{\partial x^2}=\dfrac{\partial^2 \Lambda[U]}{\partial t^2}=\dfrac{\partial \Lambda[U]}{\partial
U}=0.
\]

The four obvious linearly independent solutions of these determining equations can be obtained using the command
\[
\verb"multipliers_sol:=pdsolve(simplified_eqs[Solved]);"
\]
and are given by
\[
\Lambda[U] = C_1 + C_2 x + C_3 t + C_4 tx,
\]
where $C_1,\ldots,C_4$ are arbitrary constants.

Finally, we apply the direct method of flux computation (Section \ref{subsec:meth:direct}), using the command
\[
\verb|gem_get_CL_fluxes(Lam_sol, method="Direct");|
\]
which generates the divergence expressions of the four conservation laws
\eqref{eq:ch1:secCL:FluxesHowTo:NLWave:Cl1}, \eqref{eq:ch1:secCL:FluxesHowTo:NLWave:Cl2},
\eqref{eq:ch1:secCL:FluxesHowTo:NLWave:Cl3} and \eqref{eq:ch1:secCL:FluxesHowTo:NLWave:Cl4}.

\subsection{The first four conservation laws of the KdV equation}

As it is well-known, the KdV equation \eqref{eq:ch1:sec13:KdV:secCL} has an infinite countable sequence of local
conservation laws of increasing orders. Here we compute the first four of these conservation laws, by limiting the
multiplier dependence, as follows:
\[
\Lambda[U]=\Lambda(t,x,U, U_x,  U_{xx}).
\]

The variables and the PDE \eqref{eq:ch1:sec13:KdV:secCL} are defined using the commands
\beq\label{eq:ch5:sec_symb_pack:CLs:KdV:comm1}
\barr
\verb"with(GeM):"\\
\verb"gem_decl_vars(indeps=[x,t], deps=[U(x,t)]);"\\
\verb"gem_decl_eqs([diff(U(x,t),t)=U(x,t)*diff(U(x,t),x)+diff(U(x,t),x,x,x)],"  \\
\qquad \verb"solve_for=[diff(U(x,t),t)]);"
\earr
\eeq

The set of determining equations for the multipliers is obtained and simplified using the routines
\beq \label{eq:ch5:sec_symb_pack:CLs:KdV:comm3}
\barr
\verb"det_eqs:=gem_conslaw_det_eqs([x,t, U(x,t), diff(U(x,t),x),"\\
\qquad \verb"diff(U(x,t),x,x)]):"\\
\verb"CL_multipliers:=gem_conslaw_multipliers();"\\
\verb"simplified_eqs:=DEtools[rifsimp](det_eqs, " \verb"CL_multipliers, mindim=1);"
\earr
\eeq
and is solved using the \verb|Maple| command
\[
\verb"multipliers_sol:=pdsolve(simplified_eqs[Solved]);"
\]
to yield the four local multipliers
\beq \label{eq:KdV:4mul}
\Lambda[U] = C_1 + C_2U + C_3(x-tU)+C_4\Big(\tfrac{1}{2}U^2+U_{xx}\Big),
\eeq
where $C_1,\ldots,C_4$ are arbitrary constants.

We now reconstruct the fluxes.

\bigskip\noindent\textbf{1. The first homotopy formula.} Using the command
\beq\label{eq:use_homo_1}
\verb|gem_get_CL_fluxes(Lam_sol, method="Homotopy1");|
\eeq
we call the routine for the first homotopy method described in Section \ref{subsec:meth:herem}. It yields the conservation
laws
\beq \label{eq:KdV:4CL:Hereman}
\barr
\sg{D}_t(u) + \sg{D}_x\Big(\frac{1}{2}u^2+u_{xx}\Big)=0,\\[2ex]
\sg{D}_t\Big(\frac{1}{2}u^2\Big) + \sg{D}_x\Big(\frac{1}{3}u^3-\frac{1}{2}u_x^2 +uu_{xx}\Big)=0,\\[2ex]
\sg{D}_t\Big(xu-\frac{1}{2}tu^2\Big) + \sg{D}_x\Big(-\frac{1}{3}tu^3+\frac{1}{2}(xu^2+tu_x^2)-u_x +(x-tu)u_{xx}\Big)=0,\\[2ex]
\sg{D}_t\Big(\frac{1}{6}u^3+\frac{1}{2}uu_{xx}\Big) +
\sg{D}_x\Big(\frac{1}{8}u^4+\frac{1}{2}(u_tu_x-uu_{tx}+u^2u_{xx}+u_{xx}^2)\Big)=0.
\earr
\eeq

\bigskip\noindent\textbf{2. The second homotopy formula.} The corresponding routine for the method described in Section
\ref{subsec:meth:BA} is called as follows
\[
\verb|gem_get_CL_fluxes(Lam_sol, method="Homotopy2");|
\]
and yields the four divergence expressions in the same form as in \eqref{eq:KdV:4CL:Hereman}. By default, the arbitrary function
$\widetilde{U}=0$ is used. One can override this setting and use, e.g., $\widetilde{U}=x$, by using an option, as follows:
\[
\verb|gem_get_CL_fluxes(Lam_sol, method="Homotopy2", arb_func_Homotopy2={U=x});|
\]
which yield four conservation laws equivalent to the ones given in \eqref{eq:KdV:4CL:Hereman}, but having somewhat more
complicated fluxes:
\beq \label{eq:KdV:4CL:BA_x}
\barr
\sg{D}_t(x(t-1)+u) + \sg{D}_x\Big(\frac{1}{2}(u^2-x^2)+u_{xx}\Big)=0,\\[2ex]
\sg{D}_t\Big(tx^2+\tfrac{1}{2}(u^2-x^2)\Big) + \sg{D}_x\Big(\tfrac{1}{2} - \tfrac{1}{3}x^3 + \tfrac{1}{3}u^3 - \tfrac{1}{2}u_x^2 + uu_{xx}\Big)=0,\\[2ex]

\sg{D}_t\Big( \frac{1}{2}x^2(3t-t^2-2) +xu-\frac{1}{2}tu^2\Big) \\[2ex]
\qquad + \sg{D}_x\Big(1+t\left(\frac{1}{3}x^3-\frac{1}{2}\right) -\frac{1}{2}x^2  -\frac{1}{3}tu^3+\frac{1}{2}(xu^2+tu_x^2)-u_x +(x-tu)u_{xx}\Big)=0,\\[2ex]

\sg{D}_t\Big(\frac{1}{6}x^3(3t-1)+ \frac{1}{6}u^3+\frac{1}{2}(u-x)u_{xx}\Big)\\[2ex]
\qquad + \sg{D}_x\Big(\frac{1}{8}(u^4-x^4) +\frac{1}{2}(u_t(u_x-1)-(u-x)u_{tx}+u^2u_{xx}+u_{xx}^2 )     \Big)=0.
\earr
\eeq

The above example illustrates that normally, to obtain simpler flux expressions, one should choose $\widetilde{U}=0$, unless
integrals in the formula \eqref{eq:ch1:secCL:FluxesHowTo:homot:flux_int} diverge.

\bigskip\noindent\textbf{3. The scaling symmetry formula.} We now reconstruct the conservation laws of the KdV equation \eqref{eq:ch1:sec13:KdV:secCL}
arising from local multipliers \eqref{eq:KdV:4mul} using the scaling symmetry method (Section \ref{subsec:meth:Anco}) with
the scaling symmetry \eqref{eq:scal_sym:KdV}. In the corresponding routine, we need to specify the point symmetry
\eqref{eq:scal_sym:KdV} in the standard form:
\[
\barr
\verb|gem_get_CL_fluxes(Lam_sol, method="Scaling",|\\
\qquad\verb| symmetry={xi_x=x,xi_t=3*t,eta_U=-2*U});|
\earr
\]
This yields the following four conservation law expressions
\beq \label{eq:KdV:4CL:BA_x}
\barr
\sg{D}_t(2u+3tu_t +xu_x) + \sg{D}_x\Big(2u^2+3tuu_t+xuu_x+4u_{xx}+3tu_{txx}+xu_{xxx}\Big)=0,\\[2ex]

\sg{D}_t\Big((2u+3tu_t+xu_x)u\Big) + \sg{D}_x\Big(u(3tu_{txx} + xu_{xxx})-3tu_xu_{tx} \\
 \qquad +3 (2u+tu_t)u_xx +   u^2( 2u+3tu_t)+u_x(xu^2-3u_x)\Big)=0,\\[2ex]

\sg{D}_t\Big( (2u+3tu_t+xu_x)(x-tu) \Big) \\
\qquad + \sg{D}_x\Big( 3tu_x^2+2u^2(x-tu) + [xu(x-tu)-3] u_x +3tu(x-tu)u_t +3t(tu_x-1)u_{tx} \\
\qquad - 3(2ut+t^2ut-x)u_{xx} +(x-ut) (xu_{xxx}+3tu_{txx}\Big)=0,\\[2ex]

\sg{D}_t\Big((2u+3tu_t+xu_x)\left(\frac{1}{2}u^2+u_{xx}\right)\Big)\\
\qquad + \sg{D}_x\Big(  \frac{1}{2}u^3(xu_x+3tu_t) +u_x^2(3tu_t+xu_x-u) +u^4 -3tuu_x u_{tx} \\
\qquad+ u[xu_x+6(tu_t+u)]u_{xx} +4u_{xx}^2 + \frac{3}{2}t(u^2+2u_{xx})u_{txx} \\
\qquad+ \left[\frac{1}{2}xu^2-3(tu_{tx}+u_x)\right]u_{xxx} +(3t u_t+xu_x+2u)u_{xxxx}\Big) =0,
\earr
\eeq
equivalent to the sets of conservation laws \eqref{eq:KdV:4CL:Hereman} and \eqref{eq:KdV:4CL:BA_x} but having significantly
more complicated forms of fluxes and densities.

\section{Conclusion} \label{sec:Concl}

The direct method of construction of conservation laws is a systematic procedure applicable to all PDEs and
PDE systems involving sufficiently smooth functions. The direct method builds on ideas present in the famous
Noether's Theorem, in particular, the use of Euler differential operators, which identically annihilate any
divergence expression. However, unlike Noether's Theorem, the direct method is not limited to PDE systems
arising from some variational principle (i.e., self-adjoint PDE systems).

Within the the direct method, one seeks multipliers, such that a linear combination of the PDEs of the system system taken with
these multipliers yields a divergence expression. Importantly, for PDE systems that can be written in a solved form
\eqref{eq:ch1:secCL:PDEsys:solved_lead}, the direct method is complete, since for such systems, all conservation laws arise as
linear combinations of equations of the system.

Determining equations for unknown multipliers are obtained from the action of Euler differential operators on
a linear combination of the PDEs of a given system, and are in many ways similar to local symmetry
determining equations.

We note that the direct method can be applied without modification to ODEs and ODE systems, yielding their integrating factors
and first integrals.

After the multipliers are computed, fluxes of the corresponding divergence expression can be reconstructed using different
approaches. Each approach can be better in one particular situation and worse or non-applicable in another (Section \ref{sec:5}).

The algorithms for the direct method and the flux reconstruction have been implemented in the symbolic
package \verb"GeM" for \verb"Maple", which was used to illustrate the four methods presented in the present
paper.

\bigskip \noindent \textbf{Acknowledgements}

The author is thankful to George Bluman and Stephen Anco for multiple discussions, to anonymous referees for
useful suggestions, and to the University of Saskatchewan and the National Sciences and Engineering Research
Council of Canada for research support.



\begin{appendix}
\zeroall
\renewcommand{\theequation}{\Alph{section}.\arabic{equation}}

\section{Proof of Theorem \ref{th:ch1:secCL:HomotopyACBA}}\label{app:a}
\begin{proof}
We start with a set of local conservation law multipliers $\{\Lambda_\sigma[U]\}_{\sigma=1}^N$ of a given PDE
system \eqref{eq:ch1:secCL:PDEsys}. These multipliers satisfy \eqref{eq:ch1:secCL:CL_mixU}.

The \emph{linearizing operator} associated with the PDE system  \eqref{eq:ch1:secCL:PDEsys} is given by
\beq\label{eq:ch1:secCL:FluxesHowTo:homot:L_R}
\barr
(\sg{L}_{R})^\sigma_\rho [U]V^\rho = \left[ {\dfrac{\partial R^\sigma [U]}{\partial U^\rho } +
\dfrac{\partial R^\sigma [U]}{\partial U_i^\rho }\sg{D}_i + \ldots + \dfrac{\partial R^\sigma [U]}{\partial
U_{i_1 \ldots i_k
}^\rho }\sg{D}_{i_1 } \ldots \sg{D}_{i_k } } \right]V^\rho,\\[2ex]
\sigma = 1,\ldots ,N,
\earr
\eeq
and its adjoint by
\beq\label{eq:ch1:secCL:FluxesHowTo:homot:Ladj_R}
\barr (\sg{L}^{*}_{R})^\sigma_{\rho}[U]W_\sigma = \dfrac{\partial R^\sigma [U]}{\partial U^\rho }W_\sigma -
\sg{D}_i \left( {\dfrac{\partial R^\sigma [U]}{\partial U_i^\rho }W_\sigma } \right) \\[2ex]
\quad + \ldots + ( - 1)^k\sg{D}_{i_1 } \ldots \sg{D}_{i_k } \left( {\dfrac{\partial R^\sigma [U]}{\partial
U_{i_1 \ldots i_k }^\rho }W_\sigma } \right),\quad \rho = 1,\ldots ,m,
\earr
\eeq
acting respectively on arbitrary functions $V=(V^1(x),\ldots,V^m(x))$ and $W=(W_1(x),\ldots,W_N(x))$.

For each multiplier $\Lambda_\sigma[U]=\Lambda_\sigma(x,U,\partial U,\ldots,\partial^l U)$, introduce the
corresponding linearizing operator
\beq\label{eq:ch1:secCL:FluxesHowTo:homot:L_Lam}
\barr
(\sg{L}_{\Lambda})_{\sigma \rho} [U]\widetilde{V}^\rho = \left[ {\dfrac{\partial \Lambda_\sigma [U]}{\partial
U^\rho } + \dfrac{\partial \Lambda_\sigma [U]}{\partial U_i^\rho }\sg{D}_i + \ldots + \dfrac{\partial
\Lambda_\sigma [U]}{\partial U_{i_1 \ldots i_l }^\rho }\sg{D}_{i_1 } \ldots \sg{D}_{i_l } }
\right]\widetilde{V}^\rho ,\\[2ex]
 \sigma = 1,\ldots ,N,
\earr
\eeq
and its adjoint
\beq\label{eq:ch1:secCL:FluxesHowTo:homot:Ladj_Lam}
\barr
(\sg{L}^{*}_\Lambda)_{\sigma\rho}[U]\widetilde{W}^\sigma = \dfrac{\partial \Lambda_\sigma [U]}{\partial
U^\rho }\widetilde{W}^\sigma - \sg{D}_i \left( {\dfrac{\partial \Lambda_\sigma [U]}{\partial U_i^\rho
}\widetilde{W}^\sigma } \right) \\[2ex]
\quad + \ldots + ( - 1)^k\sg{D}_{i_1 } \ldots \sg{D}_{i_l } \left( {\dfrac{\partial \Lambda_\sigma
[U]}{\partial U_{i_1 \ldots i_l }^\rho }\widetilde{W}^\sigma } \right),\quad  \rho = 1,\ldots ,m,
\earr
\eeq
acting respectively on arbitrary functions $\widetilde{V}=(\widetilde{V}^1(x),\ldots,\widetilde{V}^m(x))$ and
$\widetilde{W}=(\widetilde{W}_1(x),\ldots,\widetilde{W}_N(x))$.

It is straightforward to show that the operators \eqref{eq:ch1:secCL:FluxesHowTo:homot:L_R} --
\eqref{eq:ch1:secCL:FluxesHowTo:homot:Ladj_Lam} satisfy the following divergence identities:
\beq\label{eq:ch1:secCL:FluxesHowTo:homot:WLV_R} W_\sigma
(\sg{L}_{R})^\sigma_\rho [U]V^\rho - V^\rho (\sg{L}^{*}_{R})^{\sigma}_{\rho}[U]W_\sigma \equiv \sg{D}_i
S^i[V,W; R[U]],
\eeq
\beq\label{eq:ch1:secCL:FluxesHowTo:homot:WLV_Lam}
\widetilde{W}^\sigma (\sg{L}_{\Lambda})_{\sigma\rho} [U]\widetilde{V}^\rho - \widetilde{V}^\rho
(\sg{L}^{*}_{\Lambda})_{\sigma\rho}[U]\widetilde{W}^\sigma \equiv \sg{D}_i
\widetilde{S}^i[\widetilde{V},\widetilde{W}; \Lambda[U]],
\eeq
where operators $S^i$, $\widetilde{S}^i$ are defined by \eqref{eq:ch1:secCL:FluxesHowTo:homot:WLV_S} and
\eqref{eq:ch1:secCL:FluxesHowTo:homot:WLV_Stilda}.

To avoid the appearance of singular integrals, we need the following construction. Let
\beq\label{eq:ch1:secCL:FluxesHowTo:homot:U_lam}
U_{(\lambda)} \equiv U + (\lambda-1)V,
\eeq
where $U=(U^1(x),\ldots,U^m(x))$ and $V=V^1(x),\ldots,V^m(x))$ are arbitrary functions, and $\lambda$ is a
scalar parameter. Replacing $U$ by $U_{(\lambda)}$ in the conservation law identity
\eqref{eq:ch1:secCL:CL_mixU}, one has
\beq\label{eq:ch1:secCL:FluxesHowTo:homot:d_dlam}
\frac{\partial}{\partial \lambda}(\Lambda_\sigma[U_{(\lambda)}] R^\sigma[U_{(\lambda)}]) \equiv
\frac{\partial}{\partial \lambda}\sg{D}_{i}\Phi^{i}[U_{(\lambda)}] = \sg{D}_{i}\left(\frac{\partial}{\partial
\lambda}\Phi^{i}[U_{(\lambda)}]\right).
\eeq
(The last identity holds since $\lambda$ can be viewed as an additional independent variable.) The left-hand
side of \eqref{eq:ch1:secCL:FluxesHowTo:homot:d_dlam} can then be expressed in terms of the linearizing
operators \eqref{eq:ch1:secCL:FluxesHowTo:homot:L_R} and \eqref{eq:ch1:secCL:FluxesHowTo:homot:L_Lam} as
follows:
\[
\frac{\partial}{\partial \lambda}(\Lambda_\sigma[U_{(\lambda)}] R^\sigma[U_{(\lambda)}]) = \Lambda_\sigma[U_{(\lambda)}] (\sg{L}_{R})^\sigma_\rho [U_{(\lambda)}]V^\rho + R^\sigma[U_{(\lambda)}](\sg{L}_{\Lambda})_{\sigma \rho} [U_{(\lambda)}]V^\rho.
\]

From \eqref{eq:ch1:secCL:FluxesHowTo:homot:WLV_R} and \eqref{eq:ch1:secCL:FluxesHowTo:homot:WLV_Lam} with
$W_\sigma=\Lambda_\sigma[U_{(\lambda)}]$ and $\widetilde{W}^\sigma = R^\sigma[U_{(\lambda)}]$, respectively,
and using the property
\[
(\sg{L}^{*}_{R})^\sigma_{\rho}[U]\Lambda_\sigma [U] + (\sg{L}^{*}_{\Lambda})_{\sigma
\,\rho} R^\sigma [U]=0
\]
of the adjoint operators \eqref{eq:ch1:secCL:FluxesHowTo:homot:Ladj_R},
\eqref{eq:ch1:secCL:FluxesHowTo:homot:Ladj_Lam}, one can show that
\beq\label{eq:ch1:secCL:FluxesHowTo:homot:SS}
\barr
\dfrac{\partial}{\partial \lambda}(\Lambda_\sigma[U_{(\lambda)}] R^\sigma[U_{(\lambda)}]) &= V^\rho (\sg{L}^{*}_{R})^\sigma_\rho [U_{(\lambda)}] \Lambda_\sigma[U_{(\lambda)}] + \sg{D}_i S^i[V,\Lambda[U_{(\lambda)}]; R[U_{(\lambda)}]]\\
&\quad +V^\rho (\sg{L}^{*}_{\Lambda})_{\sigma\rho} [U_{(\lambda)}] R^\sigma[U_{(\lambda)}] + \sg{D}_i\widetilde{S}^i[V,R[U_{(\lambda)}]; \Lambda[U_{(\lambda)}]]\\
&= \sg{D}_i\left(S^i[V,\Lambda[U_{(\lambda)}];
R[U_{(\lambda)}]]+\widetilde{S}^i[V,R[U_{(\lambda)}]\Lambda[U_{(\lambda)}]]\right).
\earr
\eeq

Comparing \eqref{eq:ch1:secCL:FluxesHowTo:homot:d_dlam} and \eqref{eq:ch1:secCL:FluxesHowTo:homot:SS}, one
finds that
\[
\sg{D}_{i}\left(\frac{\partial}{\partial \lambda}\Phi^{i}[U_{(\lambda)}]\right) = \sg{D}_i\left(S^i[V,\Lambda[U_{(\lambda)}]; R[U_{(\lambda)}]]+\widetilde{S}^i[V,R[U_{(\lambda)}];\Lambda[U_{(\lambda)}]]\right),
\]
which implies
\beq\label{eq:ch1:secCL:FluxesHowTo:homot:before_int}
\frac{\partial}{\partial \lambda}\Phi^{i}[U_{(\lambda)}] = S^i[V,\Lambda[U_{(\lambda)}];
R[U_{(\lambda)}]]+\widetilde{S}^i[V,R[U_{(\lambda)}];\Lambda[U_{(\lambda)}]],
\eeq
up to fluxes of a trivial conservation law. Now let $V=U-\widetilde{U}$, for an arbitrary function
$\widetilde{U}=(\widetilde{U}^1(x),\ldots,\widetilde{U}^m(x))$. Then $U_{(\lambda)}=\lambda U +
(1-\lambda)\widetilde{U}$. Integrating \eqref{eq:ch1:secCL:FluxesHowTo:homot:before_int} with respect to
$\lambda$ from 0 to 1, one obtains the desired expression \eqref{eq:ch1:secCL:FluxesHowTo:homot:flux_int}.
\end{proof}

\section{Computation of fluxes of a conservation law from a symmetry/adjoint symmetry pair}\label{subsec:meth:PAIR}
\zeroall

Suppose a given PDE system \eqref{eq:ch1:secCL:PDEsys} has a local symmetry with generator (in the characteristic form)
\beq\label{eq:ch1:Symm_CL:sym_adj_CL:symm}
\hat{\sg{X}} = \hat{\eta}^\rho[u]\frac{\partial}{\partial u^\rho}.
\eeq
The symmetry components $\hat{\eta}^\rho[u]$ are solutions of the symmetry determining equations, i.e., the linearized
system
\beq\label{eq:ch1:Symm_CL:sym_adj_CL:symmsol}
 \sg{L}^\sigma_\rho [u]\hat{\eta}^\rho[u] =0,
\eeq
with $\sg{L}^\sigma_\rho\equiv (\sg{L}_{R})^\sigma_\rho$ given by \eqref{eq:ch1:secCL:FluxesHowTo:homot:L_R}.

Similarly, let $\{\omega_\sigma [u]\}_{\sigma=1}^N$ be some solution of the adjoint linearized system
\beq \label{eq:ch1:Symm_CL:sym_adj_CL:adjsymmsol}
\sg{L}^{*\;\sigma}_{\rho}[u]\omega_\sigma[u] = 0,
\eeq
with $\sg{L}^{*\;\sigma}_{\rho} \equiv (\sg{L}^{*}_{R})^\sigma_{\rho}$ given by
\eqref{eq:ch1:secCL:FluxesHowTo:homot:Ladj_R}. (Functions $\{\omega_\sigma [u]\}_{\sigma=1}^N$ are often
called \emph{adjoint symmetries}).

The linearizing operator \eqref{eq:ch1:secCL:FluxesHowTo:homot:L_R} and its adjoint
\eqref{eq:ch1:secCL:FluxesHowTo:homot:Ladj_R} are related through a well-known formula \cite{AB97}
\beq\label{eq:ch1:Symm_CL:sym_adj_CL:WLV}
W_\sigma \sg{L}^\sigma_\rho [U]V^\rho - V^\rho \sg{L}^{*\;\sigma}_{\rho}[U]W_\sigma \equiv \sg{D}_i \Psi^i[U]
\eeq
where $\Psi^i[U]$ are given by
\beq\label{eq:ch1:sec13:Bluman136:WLV_phi:APP}
\Psi^i[U] = \sum_{p=0}^{k-1} \sum_{q=0}^{k-p-1}(-1)^q \left( \sg{D}_{i_1} \ldots \sg{D}_{i_p} V^\rho\right)
\sg{D}_{j_1} \ldots \sg{D}_{j_q}\left(W_\sigma \frac{\partial R^\sigma[U]}{\partial U^\rho_{j_1 \ldots j_q i
i_1 \ldots i_p}}\right),
\eeq
where $j_1 \ldots j_q$ and $i_1 \ldots i_p$ are ordered combinations of indices such that $1\leq j_1\leq
\ldots \leq j_q\leq  i\leq  i_1\leq  \ldots \leq i_p\leq n$.

The following result is immediately obtained.

\begin{theorem}\label{th:SymPlusCL:CL}
For a given PDE system \eqref{eq:ch1:secCL:PDEsys}, let $\{\hat{\eta}^\rho[u]\}_{\rho=1}^m$ be components of its local
symmetry \eqref{eq:ch1:Symm_CL:sym_adj_CL:symm}, and $\{\omega_\sigma [u]\}_{\sigma=1}^N$ a solution of the adjoint
linearized system \eqref{eq:ch1:Symm_CL:sym_adj_CL:adjsymmsol}. Then formulas \eqref{eq:ch1:sec13:Bluman136:WLV_phi} with
$V^\rho={\eta}^\rho[U]$ and $W_\sigma=\omega_\sigma [U]$ yield fluxes of local conservation laws of the PDE system
\eqref{eq:ch1:secCL:PDEsys}.
\end{theorem}

The proof follows from noting that for the symmetry/adjoint symmetry pair $\{\hat{\eta}^\rho[u]\}_{\rho=1}^m$, $\{\omega_\sigma
[u]\}_{\sigma=1}^N$, using $V^\rho=\hat{\eta}^\rho[u]$ and $W_\sigma=\omega_\sigma [u]$ in formulas
\eqref{eq:ch1:Symm_CL:sym_adj_CL:WLV} yields $\sg{D}_i \Psi^i[u]=0$, which is a local conservation laws of the PDE system
\eqref{eq:ch1:secCL:PDEsys}.

The following remarks are important.

\begin{enumerate}
    \item Since every set of local conservation law multipliers $\{\Lambda_\sigma[U]\}_{\sigma=1}^N$ of a given PDE system
        \eqref{eq:ch1:secCL:PDEsys} satisfies the adjoint linearized equations \eqref{eq:ch1:Symm_CL:sym_adj_CL:adjsymmsol},
        it follows that in Theorem \ref{th:SymPlusCL:CL}, one can use any local symmetry and any set of local conservation
        law multipliers of a given PDE system to generate its local conservation law. (Note in some cases, it is possible
        that a so generated conservation law is trivial.)
    \item A conservation law with fluxes $\Psi^i[U]$ \eqref{eq:ch1:sec13:Bluman136:WLV_phi}, obtained
        from a symmetry/multiplier pair using Theorem \ref{th:SymPlusCL:CL}, is generally
        \emph{inequivalent} to the conservation law \eqref{eq:ch1:secCL:CL_mix} with fluxes
        $\Phi^{i}[U]$, which corresponds to the multipliers $\{\Lambda_\sigma[U]\}_{\sigma=1}^N$. One obtains the fluxes
        corresponding to the multipliers only in the case of a scaling symmetry, as discussed in Section
        \ref{subsec:meth:Anco}.
\end{enumerate}

\end{appendix}

\end{document}